# Switchable Ferroelectricity in Subnano Silicon Thin Films

Hongyu Yu[a,d,*], Shihan deng[a,d,*], Muting Xie[a,d], Yuwen Zhang[b], Xizhi Shi[b], Jianxin Zhong[c,b], Chaoyu He[b,#], Hongjun Xiang[a,d,†]

[a]*Key Laboratory of Computational Physical Sciences (Ministry of Education), Institute of Computational Physical Sciences, State Key Laboratory of Surface Physics, and Department of Physics, Fudan University, Shanghai 200433, China*

[b]*School of Physics and Optoelectronics, Xiangtan University, Xiangtan 411105, China*

[c]*Center for Quantum Science and Technology, Shanghai University, Shanghai 200444, China*

[d]*Shanghai Qi Zhi Institute, Shanghai 200030, China*

*These authors contributed equally

#hechaoyu@xtu.edu.cn;

†hxiang@fudan.edu.cn:

## Abstract

Recent advancements underscore the critical need to develop ferroelectric materials compatible with silicon. We systematically explore possible ferroelectric silicon quantum films and discover a low-energy variant (hex-OR-2×2-P) with energy just 1 meV/atom above the ground state (hex-OR-2×2). Both hex-OR-2×2 and hex-OR-2×2-P are confirmed to be dynamically and mechanically stable semiconductors with indirect gaps of 1.323 eV and 1.311 eV, respectively. The ferroelectric hex-OR-2×2-P exhibits remarkable in-plane spontaneous polarization up to 120 Pc/m and is protected by a potential barrier (13.33 meV/atom) from spontaneously transitioning to hex-OR-2×2. To simulate the switching ferroelectricity in electric fields of the single-element silicon bilayer, we develop a method that simultaneously learns interatomic potentials and Born effective charges (BEC) in a single equivariant model with a physically informed loss. Our method demonstrates good performance on several ferroelectrics. Simulations of hex-OR-2×2-P silicon suggest a depolarization temperature of approximately 300 K and a coercive field of about 0.05 V/Å. These results indicate that silicon-based ferroelectric devices are feasible, and the ground state phase of the silicon bilayer (hex-OR-2×2) is an ideal system. Our findings highlight the promise of pure silicon ferroelectric materials for future experimental synthesis and applications in memory devices, sensors, and energy converters.

**Main text**

Two-dimensional (2D) ferroelectric materials has garnered significant attention, owing to their promising applications in sensors, actuators, and energy devices [1–5]. Discoveries in materials like $CuInP_2S_6$ [6], SnTe [7,8], and $In_2Se_3$ [9] have expanded the range of materials for low-dimensional memory devices [1,10]. Particularly compelling is the potential to integrate them with silicon chips, a development that could revolutionize high-performance, low-power memory, and electronic devices. Hafnium oxide ($HfO_2$)-based materials have garnered interest due to their stable ferroelectric properties at thicknesses less than 10 nanometers and compatibility with existing silicon technology [11,12].Unfortunately, $HfO_2$-based ferroelectric materials are plagued by challenges like fatigue, imprinting, and difficulties with surface manipulation, which adversely affect device performance and reliability [13]. Recent researches suggest potential ferroelectricity in single-element 2D bismuth [14] and silicon [15,16]. These findings challenge conventional understanding and open new avenues [14,17–20]. However, researches about silicon-based ferroelectricity leave critical aspects such as the Curie temperature ($T_c$) and the feasibility of polarization reversal under electric fields unresolved. Addressing these issues is pivotal for realizing the full potential of 2D ferroelectrics integrated with silicon substrates [21,22].

To investigate the ferroelectric behavior of silicon bilayer, *ab initio* molecular simulations can be a powerful tool. However, large-scale first-principles simulations require extensive computational resources. Recently developed machine learning potentials offer a promising alternative for such simulations [23,24]. As for simulations under electric field, most machine learning potentials typically treat Born effective charges (BECs) as constants or use point charge models. However, these approaches with fixed zero charge model are not suitable for single-element materials like silicon bilayer. Current potentials are limited to energy predictions and do not account for BECs. Therefore, developing a novel neural network that dynamically predicts potential energy surface (PES) and BECs based on atomic structure is thus critical.

In this Letter, we report the discovery of a metastable ferroelectric phase in silicon bilayer with group and graph theory (RG2) [25–27]. Possible silicon quantum films of reconstructed silicon bilayer are systematically explored in a high-throughput manner, revealing numerous ferroelectric phases. A noteworthy semiconducting ferroelectric phase, hex-OR-2×2-P, exhibits remarkable energetic stability, impressive in-plane spontaneous polarization and structural features similar to the ground state hex-OR-2×2. The transition barrier from the non-ferroelectric hex-OR-2×2 to hex-OR-2×2-P indicates that the metastable phase can be effectively employed for information storage. Furthermore, we have developed a dielectric response equivariant atomistic model (DREAM), which simultaneously learns MLFF and BEC and validated by several ferroelectrics. This model enables efficient and accurate simulations of thermodynamic properties and atomic behavior in electric fields, even for single-element materials. Molecular dynamics (MD) simulations based on DREAM indicate a depolarization room temperature and small coercive field for hex-OR-2×2-P silicon. Additionally, MD results demonstrate that information can be written using either metastable polarized states (hex-OR-2×2-P) or non-ferroelectric ground states (hex-OR-2×2) as the starting point.

Prediction of silicon quantum films derived from silicon bilayer has been widely studied [28–33]. Using the random method based on graph and group theory implemented in RG2, known for generating crystal structures with specific geometry features [26,27,31,34–37], we produce a variety of silicon bilayer configurations with detailed search information provided in the supplementary materials (SM) and Fig. S1-S3. The relative stabilities of these 2D silicon quantum films are assessed by plotting their total energies against layer thickness (Fig. 1(a)), with the silicene single layer as the zero reference. All reconstructions are energetically more stable than silicene, with many having lower energies than silicon bilayer (AB-1 [30]) and even silicon tri-layer (ABC). This aligns well with previous calculations [29,31], confirming the reliability of our results. Hex-OR-2×2 is identified as the energy ground state structure between silicon quantum films whose layer thickness below 5.7 Å [29]. While the ground state remains the nonpolar hex-OR-2×2, RG2 uncovers several low-energy polar configurations. Hex-OR-2×2-P, as the polarized state with the lowest energy for layer thickness below 5.7 Å, closely matches the energy and thickness of the ground state hex-OR-2×2 [29], attracting our further research.

As shown in Fig. 1(c), the structures of hex-OR-2×2 and hex-OR-2×2-P are similar, with comparable physical stability (only 1 meV/atom higher). Their excellent phase stabilities are confirmed by several calculations including phonon band structures, elastic constants and ab initio molecular dynamics with details in SM and Fig. S4, S5. It indicate that both hex-OR-2×2 and hex-OR-2×2-P can maintain their structures and properties for real application under certain conditions. Meanwhile, when bulk silicon is thinned to two layers, it spontaneously transforms into a hex-OR-2×2 or hex-OR-2×2-P, as shown in Fig. S6. The electronic band structures calculated from the HSE06 functional show that both hex-OR-2×2 and hex-OR-2×2-P are indirect band gap semiconductors with energy gaps of 1.323 eV and 1.311 eV, respectively (Fig. S7). As shown in Fig. 1(c), there is one pair of 3-coordinated Si atoms per cell at each surface in hex-OR-2×2 and hex-OR-2×2-P, forming dangling bonds. As discussed in previous literatures [15], the unsaturated Si dimers are usually titled to reduce Coulomb repulsion and lower system energy. Furthermore, the two Si atoms in each dimer locate in different geometric environments, reflecting distinct electronegativities, leading to charge transfer from inner to outer atoms and generating a dipole moment [15,16]. These dipoles, as confirmed in Bi and Sb monolayers with tilted surface dimers [38–40], will induces spontaneous polarization at the surface under consistent alignment. In hex-OR-2×2-P, the superimposed and elongated polarization between surfaces make the system ferroelectric, while in hex-OR-2×2, the polarization cancels out, making it antiferroelectric. The value of BECs of dimer atoms of structure I and II are large instead of zero and shown in Fig. 1(c). The BECs vary significantly for atoms of different structures. The corresponding $Z_{11}$ component change the sign when the polarization changes, indicating the charge transfer. The magnitude of spontaneous polarization plays a crucial role in determining the readability of information and the sensitivity of sensors. Our first-principles calculations confirmed that the ferroelectric hex-OR-2×2-P processes spontaneous in-plane polarization up to 120 Pc/m, significantly greater than those of most previously proposed 2D FEMs such as MX2 [41] and Bi-monolayer [14,19]. This value is about 3-times larger than the reconstructed Si-001 surface [16] and is comparable to the reconstructed Pentasilicene [15]. Please note that Pentasilicene is also a hypothetical configuration reconstructed from Si-001 surface at critical thickness and possess relatively higher energy. We have also confirmed that all possible reconstructed Si-001 surface with thickness comparable to hex-OR-2×2 and hex-OR-2×2-P are less stable than them [Fig. S8]. These results suggest that our new proposed hex-OR-2×2-P is highly expected to be synthesized in future experiment as a pure silicon ferroelectric.

The energy surface and phase transition pathways between polarization states are commonly of interest in studying ferroelectric properties. As indicated in Fig. 1(b), the non-ferroelectric hex-OR-2×2 and ferroelectric hex-OR-2×2-P can be characterized by the angles ($\theta_1$ and $\theta_2$) of the dimer orientation on the two surfaces. The two-degree energy surface is sampled by first-principles calculations in the area of -30°<$\theta_1$,$\theta_2$<30° in the 2D $\theta_1$-$\theta_2$ space using 30×30 grid and plotted in Fig. 1(b) and Fig. S9 in projected 2D and intuitive 3D pictures, respectively. Four symmetrically distributed local minima are revealed at (21°, 21°) and (-21°, -21°) for the non-ferroelectric ground state hex-OR-2×2 with energy of -5.047 eV/atom, as well as (21°, -21°) and (-21°, 21°) for the ferroelectric metastable hex-OR-2×2-P with energy of -5.046 eV/atom. The nonpolar ground state hex-OR-2×2 structure, with two opposite polarized metastable neighbors, is ideal for metastable ferroelectrics [42]. It features two switching pathways between ferroelectric states (21°, -21°) and (-21°, 21°): 1) direct polarization switching of both surfaces, and 2) stepwise switching via the nonpolar ground state. The single-side switching barrier is 13.36 meV/atom, significantly lower than the double-side barrier of 25.61 meV/atom as shown in Fig. 1(b) and Fig. S10. NEB-based [43] results (see Fig. S11-S12) confirm the energy barriers and polarization along the paths. The energy barrier between states suggests that hex-OR-2×2-P will not switch to hex-OR-2×2 even though hex-OR-2×2-P is not the ground state. It can be preserved as a metastable logic state for information storage, making it a promising platform for metastable ferroelectric [42]. Meanwhile, the BEC tensor analysis of $Z_{13}$ component of dimer atoms within hex-OR-2×2-P [see Fig. 1(c)] reveals that the application of an electric field along the x-axis subjects the skewed dimer atoms to opposing forces along the z-axis. This interaction prompts a possible directional shift among the dimers under electric fields, resulting in a switch of polarization.

Then we investigate the atomic behavior of the metastable ferroelectric phase hex-OR-2×2-P at room temperature and simulate its complex interactions in electric fields. As single-element materials, it's hard to assign each atom a predefined value of BECs or charge, as these values vary significantly and can change from positive to negative during simulations. To dynamically predict BECs based on the atomic structure, we develop the dielectric response equivariant atomistic model (DREAM), which integrates machine learning force fields [44,45] and BEC predictions into a single multitask model as illustrated in Fig. 2(a). It is based on equivariant neural networks (ENN), which process non-invariant geometric inputs such as displacement vectors while maintaining symmetry, resulting in internal features compliant with the 3D Euclidean group [46]. Most ENN potentials use geometric tensors, which are ultimately abandoned at last, to predict energy scalars. In contrast, DREAM utilizes the tensor information to predict BEC, which can be used to simulate the dielectric response in an electric field. BEC is a rank-2 non-symmetric tensor and can be decomposed onto a direct sum of irreducible representations (irreps) including $(l = 0, p = +1)$, $(l = 1, p = +1)$ and $(l = 2, p = +1)$. The tensor information from the output of the E(3) equivariant network is used to predict the BEC of each atom. For insulators, the whole system should remain electrically neutral. As a result, the summation of the atomic BECs for the whole system should be zero. During training, not only energy, forces and stress are included in the loss function, but also the BECs and the summation of the BECs from each atom. We propose the physical-informed loss function as:

$$L = L_E + L_F + L_{Stress} + L_{BEC} + L_{BEC_0}$$

where $L_E, L_F, L_{Stress}, L_{BEC}, L_{BEC_0}$ denote the loss function for energy, force, stress, BEC, and neutral system BEC, respectively. We choose Allegro [47], a strictly local ENN interatomic potential architecture, as the backbone of our proposed DREAM model. It is utilized to create the potential with the dielectric response and perform large-scale molecular dynamics in parallel. DREAM is equipped to scale to large system sizes with ease thanks to the strict locality of its geometric representations in Allegro with small additional computational costs to predict BECs. The atomic force under electric fields can be written as:

$$F_{i,\beta} = -\frac{\partial E(u)}{\partial u_{i,\beta}} + \sum_{\alpha} Z_{i,\alpha,\beta}(u)\mathcal{E}_\alpha$$

where $u$ notes atomic displacements, $\mathcal{E}$ notes electric fields, $E(u)$ represents the system energy without electric fields, $F_{i,\beta}$ notes force vectors of $i$ th atom in $\beta = \{x, y, z\}$ direction, $Z_{i,\alpha,\beta}$ notes BEC of $i$ th atom, $Z_{i,\alpha,\beta} = \frac{\partial P_\alpha}{\partial u_{i,\beta}}$, and $P_\alpha$ notes the polarization in $\alpha = \{x, y, z\}$ direction. In LAMMPS, simulation in electric fields based on BECs is implemented. Details of DREAM-Allegro and simulations are provided in the SM.

To validate our method, we generate datasets for $LiNbO_3$, $ZnHfN_2$ [48,49] and silicon bilayer and the results are shown in Table 1, demonstrating good performance. This furnishes a reliable model for modeling under electric fields. Furthermore, we validated the model by calculating the energy and the $Z_{13}^*$ component of BECs for silicon bilayer along the path 1 and path 2 as seen in Fig. 2(b) (see full component comparison in Fig. S13), which align well with the DFT results. Notably, even though the sample's structure is primarily concentrated around path2 shown in Fig 1(d), the model also accurately predicts the structures along path1. This indicates the model's strong generalization capability.

| System | Energy MAE (meV/atom) | Force MAE (meV/Å) | Stress MAE (meV/Å$^3$) | Born Effective Charges MAE (e) |
|---|---|---|---|---|
| silicon bilayer | 0.156 | 17.6 | 0.092 | 0.0578 |
| $LiNbO_3$ | 0.294 | 18.2 | 0.312 | 0.0658 |
| $ZnHfN_2$ | 0.198 | 10.3 | 0.411 | 0.0464 |

**Table 1**. DREAM performance on different ferroelectrics including bilayer silicon, $LiNbO_3$ and $ZnHfN_2$, as measured by mean absolute error.

The potential of hex-OR-2×2 and hex-OR-2×2-P for applications in ferroelectric storage devices (FESDs) raises our interest regarding their experimental feasibility. We therefore conducted a study on their thermodynamic stability using molecular dynamics (MD) simulations based on our neural network model. To better describe the variation in structural characteristics with temperature, we defined antiferroelectric and ferroelectric orders for hex-OR-2×2 and hex-OR-2×2-P, respectively, as $\text{order}_{\text{AFE}} = \langle \overline{\theta_1} + \overline{\theta_2} \rangle$ and $\text{order}_{\text{FE}} = \langle \overline{\theta_1} - \overline{\theta_2} \rangle$, where $\overline{\theta}$ represents the average angle of different dimers and $\langle ... \rangle$ denotes the ensemble average at various temperatures. The variations of $\text{order}_{\text{AFE}}$ and $\text{order}_{\text{FE}}$ with temperature, during the heating process from 50K to 500K, are illustrated in Fig. 3(a). Below 300K, the system retains the initial hex-OR-2×2 or hex-OR-2×2-P configuration within 50 ps after reaching thermal equilibrium, and after 300K, the direction of dimers gradually becomes disordered, resulting in the order gradually approaching 0. However, during the cooling process, the system will not return to the ground state hex-OR-2×2, but will stay in a metastable state of a random dimer distribution, as shown in Fig. 3(c). The simulation results show that both hex-OR-2×2 and hex-OR-2×2-P remain stable at room temperature and can reliably store information under power failure conditions. Further DFT calculations under 5% compressive stress reveal that the energy difference between hex-OR-2×2-P and the flat transition structure II in Fig. 1(c) is 31 meV/atom—higher than the value without stress of 23 meV/atom with detailed in Table. S1. This indicates that compressive stress could further increase the Curie temperature to about 400 K.

To confirm the information writing capability, we conducted further MD simulations to study the system's response to an external electric field. The hysteresis loop at room temperature (300K) starting from hex-OR-2×2 [see Fig. 3(b)] shows that hex-OR-2×2 can be switched to hex-OR-2×2-P in electric field. The critical electric field ($E_c$) for polarization switching of hex-OR-2×2-P is approximately 0.05 V/Å, which is lower than that of many other materials. Additionally, the hysteresis loop is maintained over four cycles of varying electric fields, indicating strong resistance to fatigue. Notably, if the simulation initiates with a multiple domains (Fig. 3(c)) formed in a cooling process, the hysteresis curve displays complete ferroelectric behavior as seen in Fig. 3(d). This result suggests that the initial state only affects the early stages of polarization establishment. This conclusion is also applicable to the case of head-to-head domain wall [see Fig. S14].

In conclusion, the group and graph theory (RG2) is employed to explore elementary ferroelectricity in reconstructed silicon bilayers. We found many ferroelectric candidates including an asymmetric one (hex-OR-2×2-P) with energy just 1 meV/atom higher than the antiferroelectric ground state of hex-OR-2×2. The ferroelectric hex-OR-2×2-P was confirmed to be dynamically and mechanically stable semiconductors with remarkable in-plane spontaneous polarization up to 120 Pc/m. With the hex-OR-2×2 as intermediate state, the single-side switching barrier between the two logic states in hex-OR-2×2-P is evaluated to be 13.33 meV/atom. Furthermore, the dielectric response equivariant atomistic model is developed to predict the potential energy surface and Born effective charges simultaneously and simulate electric-induced behaviors, validated by several ferroelectrics. With the simulations based on the trained model of silicon bilayer, the potential of hex-OR-2×2-P for ferroelectric storage devices was evaluated with stable ferroelectric orders below 300K, showing reliable information storage. The polarization switching field of 0.05 V/Å in room temperature, strong fatigue resistance, and complete ferroelectric behavior confirmed their information writing capability.

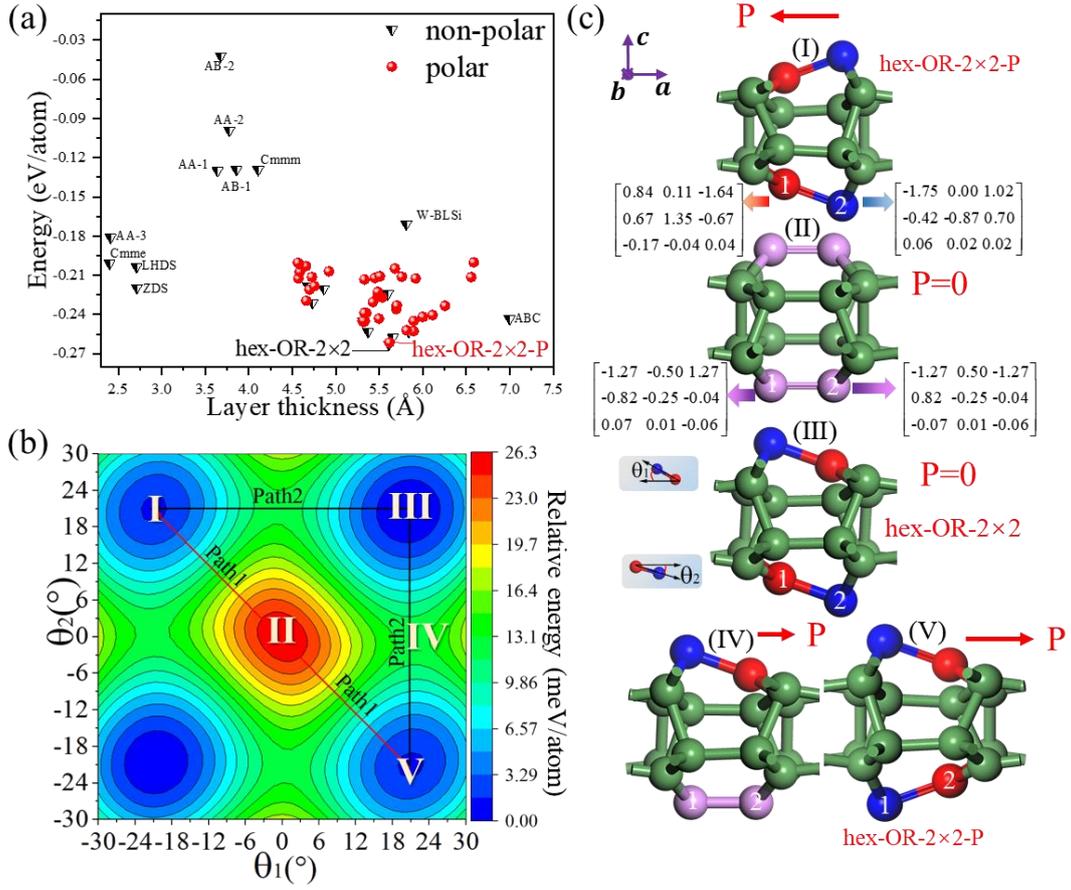

**Fig 1.** (a) The results on potential silicon bilayers. Polar and non-polar new configurations are represented by red solid balls and black-white alternating inverted triangles, respectively. (b)The two-degree-of-freedom model for describing hex-OR-2×2 and hex-OR-2×2-P, as well as the corresponding structures and 2D potential energy surface E=E($\theta_1$, $\theta_2$). (c) The structures on (b) and BEC tensor of represented dimer and inner atoms of hex-OR-2×2-P and structure II. The silicon atoms with positive, nagetive or neutral charge are shown in blue, red or pink, respectively.

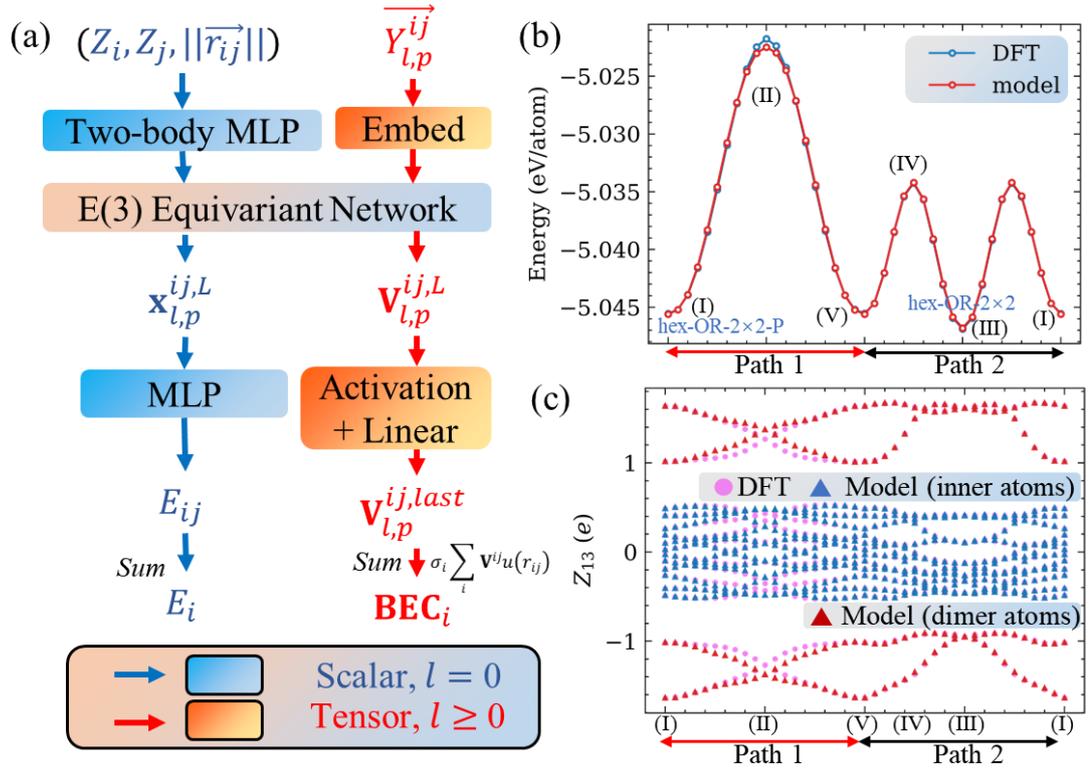

**Fig 2.** (a) shows the DREAM framework. Scalar and tensor information from the output of E(3) equivariant network, represented by blue and red, are used to predict energy and BECs, respectively. (b) and (c) shows the energy and $Z_{31}$ component of BECs comparison between DFT and trained model of silicon bilayers for the structures along the path.

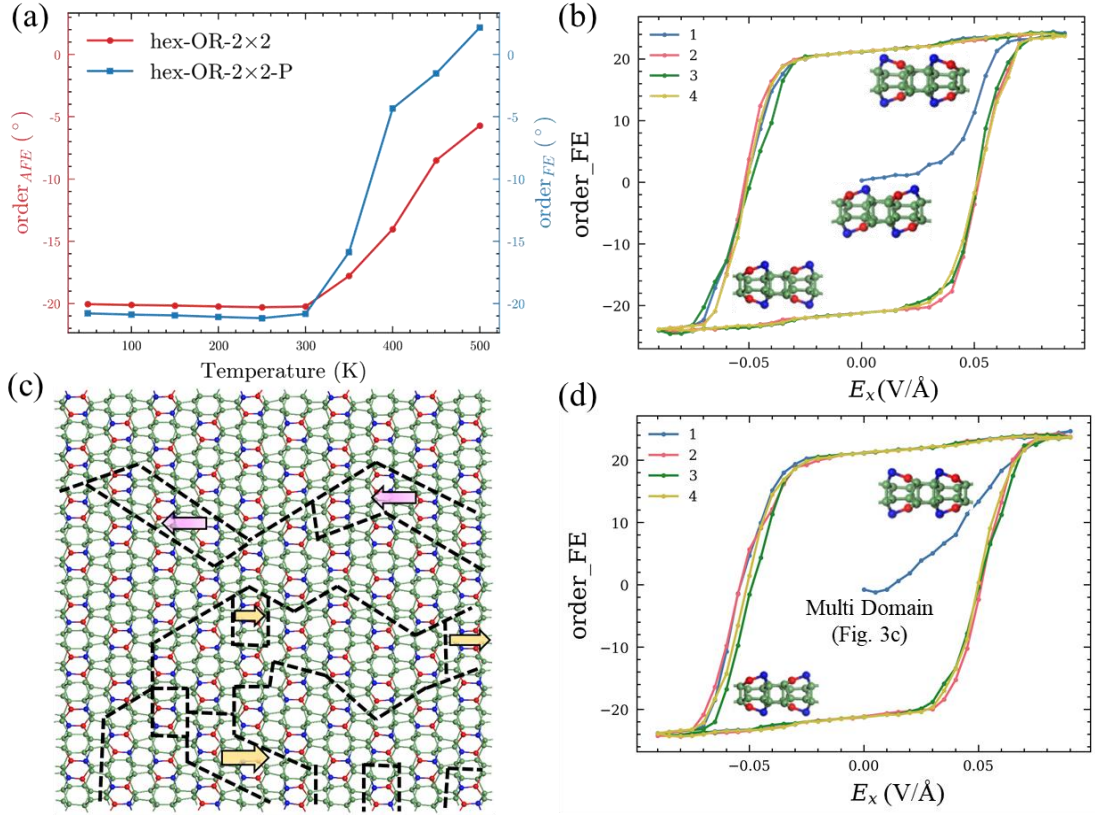

**Fig 3.** (a) shows the thermodynamic stability of hex-OR-2×2-P and hex-OR-2×2 using molecular dynamics with trained model under different temperature. (b) shows the hysteresis loop of hex-OR-2×2 under 300 K by electric-field induced molecular dynamics with trained model. The coercive electric field is about 0.05 V/Å. (c) shows the mutil-domain structure from a cooling simulation, which serves as the initial state of (d) the hysteresis loop simulation under 300 K.

## Acknowledgments

We acknowledge financial support from the National Natural Science Foundation of China (Grants No. 12204397 and No. 52372260), the Science Fund for Distinguished Young Scholars of Hunan Province of China (Grant No. 2021JJ10036), the National Key R&D Program of China (No. 2022YFA1402901), NSFC (No. 11825403, 11991061, 12188101, 12174060, and 12274082), the Guangdong Major Project of the Basic and Applied Basic Research (Future functional materials under extreme conditions--2021B0301030005), and Shanghai Pilot Program for Basic Research—Fudan University 21TQ1400100 (23TQ017).